\documentstyle[prl,aps,epsf,floats]{revtex}  
\newcommand\LQCD{{\Lambda_{\rm QCD}}}
\newcommand\LQCDs{{\Lambda^2_{\rm QCD}}}
\newcommand\bL{{\bar\Lambda}}

\newcommand\hq{{\hat q}}

\newcommand\vev[1]{\langle{#1}\rangle}
\newcommand\eg{{\it e.g.}}

\newcommand\beq{\begin{equation}}
\newcommand\eeq{\end{equation}}

\newcommand\vubud{{V^{\phantom{*}}_{ub}V^*_{ud}}}
\newcommand\vtbtd{{V^{\phantom{*}}_{tb}V^*_{td}}}

\begin{document}

\twocolumn[
\hsize\textwidth\columnwidth\hsize
\csname@twocolumnfalse\endcsname

\title{A Method For Extracting {\mbox {\boldmath $\cos\alpha$}}}
\author{Benjam\'{\i}n
Grinstein${}^{\dagger}$, Detlef R. Nolte${}^{\dagger}$
 and Ira Z. Rothstein${}^\ddagger$ \\[4pt]}
\address{\tighten{\it
${}^{\dagger}$Department of Physics,
University of California at San Diego, La Jolla, CA 92093 USA\\
${}^\ddagger$Physics Department, Carnegie Mellon University, Pittsburgh PA 15213 }\\[4pt]
UCSD/PTH 99--15\\[4pt] October 1999}

\maketitle

\tighten{
\begin{abstract}
We show that it is possible to extract the weak mixing angle
$\alpha$ via a measurement of the rate for  $B^\pm\to \pi^\pm
e^+e^-$. The sensitivity to $\cos \alpha$ results from 
the interference between the long and short distance
contributions. The short distance contribution can be computed, using
heavy quark symmetry, in terms of semi-leptonic form factors.
More importantly, we show that, using Ward identities
and a short distance operator product expansion,  the long
distance contribution can be calculated without recourse
to light cone wave functions when  the invariant mass of the 
lepton pair, $q^2$, is much larger than $\LQCDs$.  
We find that for $q^2\ge2~\hbox{GeV}^2$ the
branching fraction is approximately 
$1\times10^{-8}|V_{td}/0.008|^2$. 
The shape of the differential rate is
very sensitive to the value of $\cos\alpha$ at small values of
$q^2$ with $d\Gamma /dq^2$ varying up to 50\% in the interval 
$-1<\cos\alpha<1$ at $q^2=2~\mbox{GeV}^2$. The size of the
variation depends upon the ratio $V_{ub}/V_{td}$.
\end{abstract}
}
\vspace{0.2in}

]\narrowtext

\newpage

\begin{figure}
\centerline{
\epsfysize 2.5in
\epsfbox{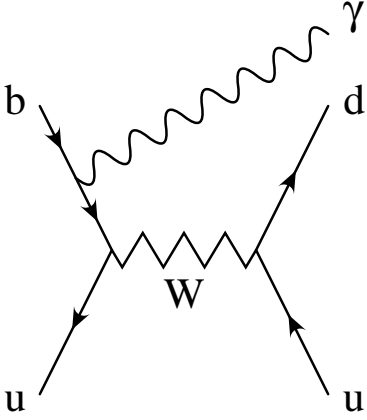}}
\vskip0.5cm
\caption{Weak annihilation diagram underlying the
decays $B\to \rho e^+e^-$ and $B\to \pi e^+e^-$. There are three other
diagrams with the photon emitted from any of the three light
quarks. Photon emission from the $W$-boson is suppressed by $G_F$.}
\label{fig:fig0}
\end{figure}

Great effort is presently being expended in attempting to understand
the origin of CP violation. It is hoped that in the next generation
of experiments we will be able to determine all the parameters
in the CKM matrix. However, the extraction of 
these parameters is, in general, hindered
by our inability to perform first principle calculations of rates due
to the non-perturbative nature of the long distance QCD effects.
A particularly nettlesome extraction
is that of the angle $\alpha$ in the unitarity triangle. 
The standard proposal for the extraction of $\alpha$  from 
$B\rightarrow \pi \pi$ is hindered
by so called penguin pollution, which can only be overcome
through a cumbersome $SU(3)$ analysis. 
Here we propose to extract this angle via a measurement of the
rate for the rare decay\footnote{ It is also
possible to use the mode $B\to \rho e^+e^-$, which will be discussed
in a separate publication\cite{us-forthcoming}.} $B\to \pi e^+e^-$.
 It is usually assumed that the rate for this process is dominated 
by the short distance
transition $b\to d e^+e^-$ except when the invariant $e^+e^-$ mass,
$q^2=(p_{e^-}+p_{e^+})^2$, is of order of charmonium resonances where
long distance contributions are important (here and below by short
distance transition we mean contributions to the amplitude that are
effectively local at distances larger than the electroweak scale
$1/M_W$). However, there is a long distance
contribution, which arises through  weak annihilation diagrams, like the one in
Fig.~\ref{fig:fig0}, 
which can contribute significantly.
The short distance amplitude is proportional to 
$V^{\phantom{*}}_{td} V^*_{tb}$ whereas
the long distance annihilation graph is proportional to
$V^{\phantom{*}}_{ud} V_{ub}^*$. 
Thus, the interference of these contributions leads
to a rate which is sensitive to the value for $\cos \alpha$ where
\beq
\alpha=\arg\left[-
\frac{V^{\phantom{*}}_{td}V_{tb}^*}{V^{\phantom{*}}_{ud}V_{ub}^*}\right].
\eeq
Even a crude measurement of $\cos\alpha$ would be of value since it
would remove a two-fold ambiguity in extractions of $\alpha$ from
$\sin2\alpha$. 
Naively, one would think that any hope of extracting $\alpha$ 
in this way is doomed by the fact that long distance contributions
are notoriously intractable. 
However, in this paper we show that 
this weak annihilation can be calculated in
an expansion in $1/m_b$ and $\alpha_s$ when the invariant mass
of the electron pair $q^2$ is larger than $\LQCDs$. Moreover, as 
will be seen below, the rate is independent of the valence, 
as well
as higher twist, wavefunctions of the pion, thereby reducing the uncertainty
in the calculation.

Let us naively estimate the relative importance of the short distance
amplitude and of the long distance weak annihilation amplitude. The
former must involve an electroweak loop, so it carries a factor of
$1/16\pi^2$, while the latter should be suppressed by wave functions
at the origin, $f_Bf_\pi/M_B^2$, where $f_B$ and $M_B$ are the
$B$-meson decay constant and mass, respectively, and $f_\pi$ is the
$\pi$-meson decay constant.\footnote{At small $q^2$ the short
distance amplitude is further suppressed by $q^2/M_B^2$.}  In
addition, the CKM factors are different. So the ratio of the long
distance to short distance amplitudes is expected to be of the order
of $|V_{ub}/V_{td}|16\pi^2 f_Bf_\pi/M_B^2\sim 0.07$. Here we have
used $f_B=170$~MeV, $f_\pi=130$~MeV, and $|V_{ub}/V_{td}|=0.6$. 
Thus, weak annihilation can easily give a correction of 10\% to the rate,
possibly larger. A more quantitative calculation is well motivated.

Let us now undertake a systematic calculation of the
weak annihilation amplitude in $B^\pm\to \pi^\pm e^+e^-$, which will
be the primary focus of this paper.
The relevant effective
Hamiltonian for the long distance, 
weak annihilation contribution to the rate for $B^-\to\pi^- e^+e^-$ is 
\beq
\label{eq:Heff}
{\cal H}'_{\rm eff}= \frac{4G_F}{\sqrt2}\,\vubud\left(
c(\mu/M_W){\cal O}+c'(\mu/M_W){\cal O}'\right),
\eeq
where 
\beq
\label{eq:Odefd}
{\cal O}=\bar u\gamma^\nu P_- b \;\;\bar d\gamma_\nu P_- u
\eeq
and
\beq
{\cal O}'=\bar u\gamma^\nu P_-T^a b\;\; 
\bar d\gamma_\nu P_- T^a u,
\eeq
$P_\pm\equiv(1\pm\gamma_5)/2$ and $T^a$ are the generators of color
gauge symmetry.
The dependence on the renormalization point $\mu$ of
the short distance coefficients $c$ and $c'$ cancels the
$\mu$-dependence of operators, so matrix elements of the
effective Hamiltonian are $\mu$-independent. At next to leading
log order, using $\Lambda^{(5)}_{\rm QCD}=225$~MeV, the coefficients
at $\mu=5$~GeV are\cite{buras} 
$c=1.02$ and $c'=-0.34$.

The $B^-\rightarrow \pi^-e^+e^-$, decay rate is given by
\begin{equation}
\label{eq:doublediffrate}
\frac{d\Gamma}{dq^2dt}=\frac1{2^8\pi^3M_B^3}
\left|\frac{e}{q^2}\ell_\mu h^{\mu} \right|^2 ,
\end{equation}
where $\ell^\mu=\bar u(p_{e^-})\gamma^\mu v(p_{e^+})$ is the leptons'
electro-magnetic current, $q\equiv p_B+p_{e^-}$ and
$t\equiv(p_D+p_{e^+})^2=(p_B-p_{e^-})^2$.  A sum over final state
lepton helicities is implicit. The non-local contribution to the
hadronic current $h$ is 
\beq
\label{eq:melem}
h^{\mu}=
\langle \pi| \int d^4x\;e^{iq\cdot x}
\; T(j^\mu_{\rm em}(x){\cal H}'_{\rm eff}(0)) |B\rangle.
\eeq
The two body kinematics with an energetic massless final state hadron
is known to factorize\cite{dugan} in the sense that soft gluon
exchange between initial and final state hadrons is suppressed by
$1/m_b$ or $\alpha_s(m_b)$. This factorization results from the
fact that the energetic outgoing light quarks form a small color
singlet object which does not couple to leading order in the ratio
$k/E_q$ (``color transparency'' \cite{bjorken}), 
where $k$ is the soft gluon momentum and $E_q$ is the
energy of the outgoing quarks which scales with $m_b$ provided $q^2$ is
not close to $q^2_{\rm max}=(M_B-m_\pi)^2$. Furthermore, by the
same reasoning, the color octet operator does not contribute to
leading order in $k/E_q$ once the final state is projected onto
the color singlet channel.
Notice that this is the {\it simplest} example of Bjorkens' color
transparency argument, since there is only one hadron in the final
state. There are no assumptions needed regarding the behavior of
wave functions. 

Returning now to our result we find that factorization leads to
\begin{eqnarray}
\label{eq:melemfactorized}
h^{(X)\mu} 
&=& 
\kappa\int d^4x\;e^{iq\cdot x} \big[\vev{ \pi| \; T(j^\mu_{\rm
 em}(x)j_\lambda(0))|0}\hbox{$\frac12$}f_B p_B^\lambda \nonumber\\ 
 & &\hspace{1cm}+\vev{\pi|j_\lambda(0)|0}
\vev{0| T(j^\mu_{\rm em}(x)J^\lambda(0)) |B}
\big]
\end{eqnarray}
where $\kappa={2\sqrt2G_F}\,\vubud c$,
$j_\lambda=\bar d\gamma_\lambda P_- u$ and $J^\lambda=\bar
u\gamma^\lambda P_- b$.

The first line in Eq.~(\ref{eq:melemfactorized}) can be computed using
an isospin Ward identity. The $B$-momentum acts as a derivative on the
$T$-ordered product which then gives the matrix element of a
commutator. It makes a contribution to 
$h^{\mu}$ of  $ -e\kappa f_\pi f_B p_B^\mu$.

It is remarkable that the result is not sensitive to the internal
structure of the pion. The first term in
Eq.~(\ref{eq:melemfactorized}) may be written using a lightcone OPE,
as a weighted integral over the pion valence wavefunction plus higher
twist corrections\cite{ruckl}. If one were to perform this
calculation, one would find that to all orders in $\alpha_s$ the
Wilson coefficients are independent of the momentum fraction carried
by the light quarks. The generality of our result allows one to see
immediately that in addition it is independent of all higher twist
wavefunctions as well.

The second line in Eq.~(\ref{eq:melemfactorized}) is also insensitive
to the lightcone wavefunction of the $B$ meson, in that it can be
computed using a short distance OPE \cite{evans-99-2,evans-99-3}
 and a heavy quark expansion (HQET) for the $b$-quark, provided $q^2\gg\LQCDs$.

To leading order we find
\beq
\label{eq:hmupi}
h^{\mu} = -{\textstyle\frac43}e\kappa f_\pi f_B p_B^\mu.
\eeq 
The coefficient of the singlet operator (\ref{eq:Odefd}) is a function
of renormalization point, $\mu$. However, once we have matched to the
HQET the $\mu$-dependence of the coefficient cancels that of the decay
constant of the $B$-meson in the HQET. This invariant combination is
the physical decay constant $f^{\phantom{\dagger}}_B$, or rather, the
leading approximation to it in an expansion in $1/m_b$.

As we will see the sensitivity to $\cos \alpha$ is greatest at small
$q^2$ where the short distance contribution is suppressed due the
rapidly falling form factors. Thus we would like to be able to trust
our results to as low a value of $q^2$ as possible.  If $k$ denotes
the momentum of the $u$-quark in the $B$-meson, the OPE is a double
expansion in $q\cdot k/q^2\sim\LQCD m_b/q^2$ and
$k^2/q^2\sim\LQCDs/q^2$. The leading correction, of order $q\cdot
k/q^2$ can be computed. The remaining corrections are of order 20\%
for $q^2\ge3.5\mbox{GeV}^2$ and can be computed in terms of matrix
elements of local operators. Including the leading correction we
find
\beq
\label{eq:hmupi2}
h^{\mu} = -{\textstyle\frac43}e\kappa f_\pi f_B p_B^\mu(1+
{\textstyle\frac23}\bL m_b/q^2),
\eeq 
where  $\bL=M_B-m_b$ is the meson
mass in HQET.  A recent extraction\cite{cleo-moments} gives
$\bL\approx330~\mbox{MeV}$.

We now combine the long and short distance contributions to the
amplitude for $B^-\to\pi^- e^+e^-$. At leading order\footnote{The
penguin operators ${\cal O}_3$--${\cal O}_6$ have small coefficients
and we neglect them. The gluonic magnetic moment operator, ${\cal
O}_8$, does not contribute to $B\to\pi\ell\ell$ at this order.}  the
short distance contribution to the amplitude is obtained from the
effective Hamiltonian\cite{gsw}
\beq {\cal H}_{\rm eff}=
\frac{4G_F}{\sqrt2}\vtbtd \sum_{j=7,9,10} 
C_j(\mu){\cal O}_j(\mu),
\eeq 
where
\begin{eqnarray}
{\cal O}_7 &=& \frac{e}{16\pi^2}\,m_b 
(\bar d\sigma^{\mu\nu}P_+b)\,F_{\mu\nu},\\
{\cal O}_9 &=& \frac{e^2}{16\pi^2}(\bar d\gamma^\mu P_-b)\,\bar
e\gamma_\mu e,\\
{\cal O}_{10} &=& \frac{e^2}{16\pi^2}(\bar d\gamma^\mu P_-b)\,\bar
e\gamma_\mu \gamma_5e.
\end{eqnarray}
At leading-log, with $\alpha_s(M_Z)=0.12$ and $m_t=175~\hbox{GeV}$,
$C_7(m_b)=0.33$ and $C_{10}(m_b)=5.3$\cite{gsw}. At next to
leading-log $C_9(m_b)=-4.3$\cite{burasmunz}. There are additional long
distance
contributions from the operators ${\cal O}$ and  ${\cal O}'$, and from
the corresponding operators with the $u$-quark replaced by a
$c$-quark, in which a photon is emitted from a $u$- or $c$-quark
loop. These contributions can be
incorporated into a shift in $C_9$,
\begin{eqnarray}
\tilde C_9=C_9&+&(c+{\textstyle\frac43}c')
g(m_c/m_b,\hq^2)\nonumber\\
&+&\frac\vubud\vtbtd 
(c+{\textstyle\frac43}c') [g(m_c/m_b,\hq^2)-g(0,\hq^2)]
\end{eqnarray}
where $\hq^2=q^2/m_b^2$ and the function $g$ is defined in Ref.~\cite{gsw}.
The short distance amplitude is given in terms of
form factors, $f_\pm$ and $h$, defined by
\begin{eqnarray}
\vev{\pi(p')|\bar d \gamma^\mu b|B(p)}&=& (p+p')^\mu f_+ + (p-p')^\mu
f_-\\
\vev{\pi(p')|\bar d \sigma^{\mu\nu} b|B(p)}&=& 2ih(p^\nu
p^{\prime\mu}-p^\mu p^{\prime\nu})
\end{eqnarray}
Including the long distance contribution to the amplitude, as calculated
above, and neglecting the mass of the electron, the rate for $B\to\pi
e^+e^-$ is
\begin{eqnarray}
\label{eq:lsrate}
\frac{d\Gamma}{dq^2}&=&|\vtbtd|^2
\frac{G_F^2\alpha^2m_B^3}{3\times2^9\pi^5}\frac{(m_B^2-q^2)^3}{m_B^6}
\bigg[|C_{10}f_+|^2+\nonumber\\
& &\hskip-1.0cm\bigg|\tilde C_9f_++2m_bC_7h-\frac{16\pi^2}3 
\frac{\vubud}{\vtbtd} 
\frac{c(m_b)f_\pi f^{\phantom{\dagger}}_B}{q^2}(1
+\frac{2\bL m_b}{3q^2})\bigg|^2\bigg]\nonumber\\
\end{eqnarray}
The contributions from the weak operator $\bar d \gamma^\mu b \bar c
\gamma_\mu c$ will be poorly described by the function $g$ when $q^2$
corresponds to the mass of a charmonium state, so we restrict our
analysis to $q^2<m^2_\psi$.

\begin{figure}
\centerline{
\epsfysize 3.5in
\epsfbox{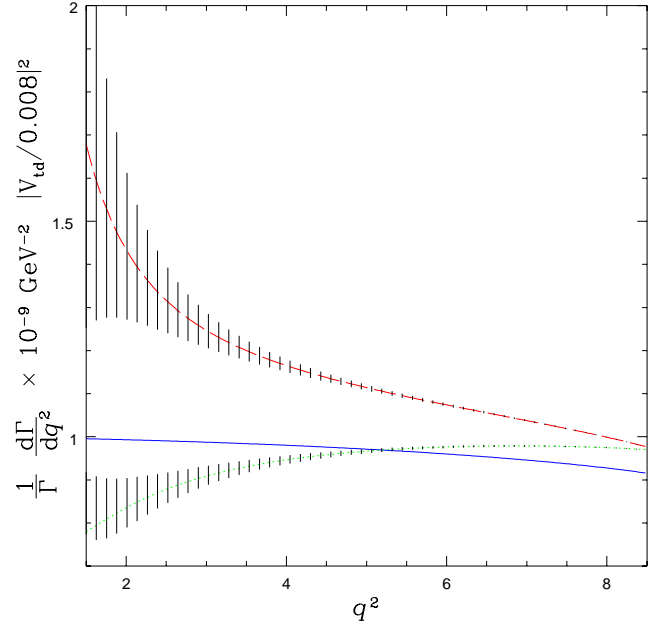}}
\vskip0.5cm
\caption{Differential branching fraction, $\tau_B\,d\Gamma/dq^2$ for
$B\to \pi e^+e^-$. The solid (blue), dashed
(red) and dotted (green) lines correspond to
$\cos\alpha=0,-1,1$, respectively. The shaded region shows the
uncertainty in our calculation. For the short distance contribution
the BK form factors of Ref.~\protect\cite{BK} have been used.}
\label{fig:lsrateBK}
\end{figure}

We see that the interference effect is largest at smaller
values of $q^2$ where the form factors are suppressed.
A part of the uncertainty in the extraction will be depend
on  our knowledge of the form factors.
The form factors $f_+$ and $h$ may be extracted from measurements
of the semi-leptonic $B$ decays and $B\rightarrow K^* \gamma$, respectively.
Alternatively, we may use heavy quark symmetry and use the relation
\beq
\label{eq:hfromHQS}
h(q^2)=\frac{\left( f_+(q^2)-f_-(q^2) \right) } {2 m_b}+O(1/m_b).
\eeq
Presently, there exists lattice QCD determinations at larger values of 
$q^2$, where there is little hadronic recoil.

Given, that presently we do not know the form factors away from largest
values of $q^2$,  for illustration purposes we will 
use the BK model (of Becirevic and
Kaidalov\cite{BK}) which satisfies the unitarity sum rules
bounds\cite{boyd} and  fits  lattice
determinations at large $q^2$:
\beq
f_+^{(\hbox{BK})}=\frac{N(1-\beta)}{(1-\tilde q^2)(1-\beta\tilde
q^2)},~~
f_0^{(\hbox{BK})}= \frac{N(1-\beta)}{1-\gamma\tilde
q^2},
\eeq
where $f_0\equiv\frac{q^2}{m_B^2-m_\pi^2}f_-+f_+$, $\tilde
q^2=q^2/M_{B^*}^2$ and the parameters are $\beta=0.54$, $\gamma=0.8$,
$N=0.6$ and $M_{B^*}=5.325$~GeV. Here we have changed the value of
$N=0.8$ given in Ref.~\cite{BK}, which is appropriate for
$m_b=2.6$~GeV, for the more realistic value of $N=0.6$. The value of
$\gamma$ is fixed by the Callan-Treiman relation
$f_0(m_B^2)=f_B/f_\pi$. The BK model does not give the form factor
$h$. We calculate $h$ using the heavy quark spin symmetry
relation~(\ref{eq:hfromHQS}).

In Fig.~\ref{fig:lsrateBK} we plot the rate of Eq.~(\ref{eq:lsrate})
as a fraction of the total width $\Gamma=\tau_B^{-1}$ as a function of
$q^2$.  We have used the coefficients $c$, $c'$ and $C_9$ at next to
leading order, and the rest at leading log order.  We have restricted
the plot to $q^2\ge1.5~\hbox{GeV}^2$ for our approximations to remain
valid, and to $q^2\le8.5~\hbox{GeV}^2$ to avoid contributions from
charmonium resonances. For illustration we have used $|\vubud|=0.004$,
$|\vtbtd|=0.008$, and $f^{\phantom{\dagger}}_B=0.17$~GeV. The solid
(blue), dashed (red) and dotted (green) lines correspond to
$\cos\alpha=0,-1,1$, respectively and the shaded region represents an
uncertainty $\pm(\bL m_b/q^2)^2$ to the correction in
Eq.~(\ref{eq:hmupi2}).  The values used for  $|\vubud/\vtbtd|$
and $f_B$ are uncertain. The 
long distance correction could be even more pronounced if they
happened to be larger.

The two limitations of our proposal are the size of the branching
fraction and the unknown form factors. We see from (\ref{eq:lsrate})
the branching fraction is sensitive to $V_{td}$ which is presently
constrained to be in the range 0.004--0.012. Thus, the total branching
fraction will vary between $10^{-7}-10^{-9}$. Such branching ratios
are most probably out of the reach of $e^+ e^-$ machines, but well
within the reach of hadronic machines given the mode.  To extract
$\cos\alpha$ we should look at the partially integrated rate for
$2\hbox{GeV}^2\le q^2\le9\hbox{GeV}^2$ which will further reduce the
rate by a factor of 2 or so. The accuracy with which we may extract
the CP violating parameters from $B^\pm\rightarrow \pi^\pm l^+l^-$ is
limited by how well we know the form factors.  While presently we do
not know their values, especially at smaller values of $q^2$, given
the plethora of data which will soon be available at the B factories
and at CDF, it is not unreasonable to believe that we will know these
form factors well enough to make our analysis useful.

We have shown that the rate for $B^\pm\to\pi^\pm e^+e^-$ is sensitive
to $\frac{\vubud}{\vtbtd}$. If the magnitude of this ratio is known
then the rate and particularly the shape of the spectrum depend
sensitively on $\cos\alpha$. Even a crude measurement of the shape
would almost certainly determine the sign of $\cos\alpha$ and remove a
two-fold ambiguity from $\sin2\alpha$. If the magnitude is not known
then a measurement of the rate and spectrum would constrain the
unitarity triangle, \eg, a region of the $(\rho,\eta)$
plane\cite{us-forthcoming}.  The analysis requires knowledge of decay
form factors.  Semi-leptonic decay spectra will determine the
combination $|V_{ub}f_+|$. Such a measurement can be incorporated in
our analysis\cite{us-forthcoming} and constrain the unitarity
triangle in a meaningful way even if separate knowledge of $\mid V_{ub}\mid$
and $f_+$ is lacking.

\bigskip

{\it Acknowledgments} The authors benefited from conversations with
Jim Russ. B.G. and D.R.N. are  supported by the Department of
Energy under contract No.\ DOE-FG03-97ER40546. I.R. is supported
by the Department of
Energy under contract No.\ DOE-ER-40682-143.

\tighten

\end{document}